\documentclass[reprint,
superscriptaddress,
 amsmath,amssymb,
 aps,
pra,
]{revtex4-2}

\usepackage{graphicx}
\usepackage{dcolumn}
\usepackage{bm}
\usepackage{tabularx}
\usepackage[shortcuts]{extdash} 
\usepackage{xcolor} 
\usepackage{amsmath}
\usepackage{siunitx}
\usepackage{enumitem}

\begin{document}

\title{Inverse Design of Integrated Photonic Components for Visible\-/Light Applications}

\author{Christoph Stockinger}
\thanks{These authors contributed equally to this work.}
\affiliation{Institute of Physics, University of Graz, NAWI Graz, Universitätsplatz 5, 8010, Graz, Austria}
\affiliation{Christian Doppler Laboratory for Structured Matter Based Sensing, Universitätsplatz 5, 8010, Graz, Austria}

\author{Marko {\v S}imi{\'c}}
\thanks{These authors contributed equally to this work.}
\affiliation{Institute of Physics, University of Graz, NAWI Graz, Universitätsplatz 5, 8010, Graz, Austria}
\affiliation{Christian Doppler Laboratory for Structured Matter Based Sensing, Universitätsplatz 5, 8010, Graz, Austria}

\author{John O. Adeniran }
\affiliation{University of Glasgow, Rankine Building, Oakfield Avenue, Glasgow G12 8LT, UK}

\author{Natale G. Pruiti}
\affiliation{University of Glasgow, Rankine Building, Oakfield Avenue, Glasgow G12 8LT, UK}

\author{Marc Sorel}
\affiliation{University of Glasgow, Rankine Building, Oakfield Avenue, Glasgow G12 8LT, UK}
\affiliation{Sant’Anna School of Advanced Studies, Via Moruzzi 1, 56124 Pisa, Italy}

\author{Peter Banzer}
\email{peter.banzer@uni-graz.at}
\affiliation{Institute of Physics, University of Graz, NAWI Graz, Universitätsplatz 5, 8010, Graz, Austria}
\affiliation{Christian Doppler Laboratory for Structured Matter Based Sensing, Universitätsplatz 5, 8010, Graz, Austria}
\email{peter.banzer@uni-graz.at}


\begin{abstract}
Photonic integrated circuits (PICs) operating in the visible spectral range are crucial for quantum technologies, optical sensing, and nonlinear optics applications. However, their development is hampered by limited integration density and operational bandwidth, linked to low refractive index contrast and restricted parameter space in intuition-based PIC design. We address these bottlenecks with an adjoint-based, fabrication-aware inverse design workflow tailored for visible-spectrum photonics. By expanding the capabilities of FDTDX, an open-source, GPU-accelerated FDTD solver, we provide a versatile memory and runtime efficient inverse design platform. Demonstrating this approach on a silicon nitride ($\mathrm{Si_3N_4}$) platform, we design, fabricate, and experimentally validate ultra-compact, broadband components for light routing, multiplexing, and polarization control. This work provides a scalable framework for high-density, high-performance visible-light PICs.
\end{abstract}

\maketitle

%
\begin{figure*}[t]
	\centering
	\includegraphics[width = 1\textwidth]{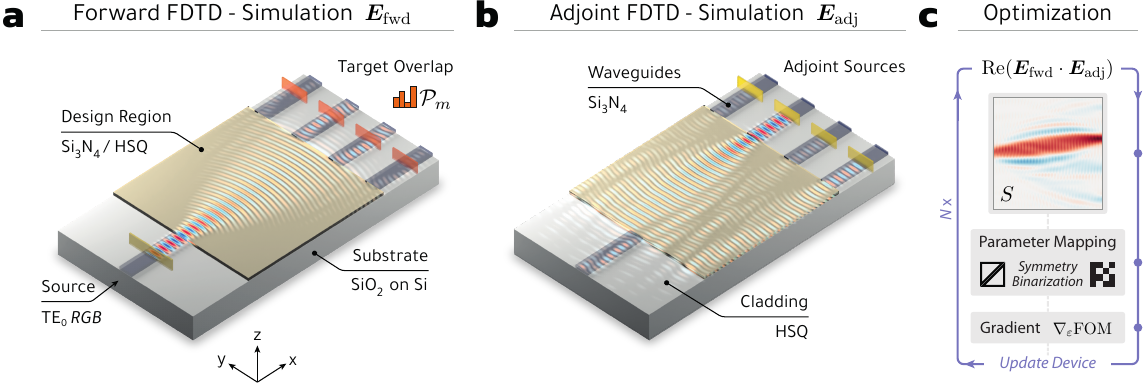}
	\caption{Schematics of adjoint-based optimization using a power splitter as an example, implemented on a $\mathrm{Si_3N_4}$ platform. (a) In the forward simulation, an input mode is launched via a waveguide and propagates across the device geometry. Monitors at the output ports of the device record the modal complex amplitude, enabling the calculation of the modal overlap with a desired output mode and its corresponding power $\mathcal{P}_m$. (b) In the adjoint simulation, objective-weighted modes are launched backwards from the output ports, generating the adjoint fields in the \textit{Design Region}. (c) The forward and adjoint fields are combined in the design region, and from their spatial overlap one infers information about the sensitivity $S$ of the objective to local material changes. Applying material parameterization (e.g., smoothing, binarization, symmetry) projects the sensitivity back onto the optimization parameters, yielding the gradient used to update the design, thus allowing for the optimization of the objective. This optimization process is repeated until the target performance or maximum number of iterations, $N$, is reached.}
	\label{fig:schematicsadjoint}
\end{figure*}
\section{\label{sec:intro}Introduction}

Photonic integrated circuits (PICs) promise exceptional potential for miniaturization, cost-effective fabrication at large scales, and on-chip integration, positioning them at the forefront of next-generation photonic technologies.
In particular, silicon-based photonic circuits have reached a high level of maturity~\cite{Thomson_2016}, enabling the realization of highly complex systems for applications such as photonic computing~\cite{Shastri2021,Pai2023}, light detection and ranging~\cite{Milanizadeh2022, Butow2024, Butow2023,Settembrini2026}, and other emerging fields.

In contrast to the well-established development of silicon photonics in the telecommunications wavelength range~\cite{schul2026inverse}, integrated photonics in the visible spectral range is still an emerging field, where simultaneously achieving low-loss propagation, high integration density, and scalable fabrication remains challenging \cite{PORCEL2019299, https://doi.org/10.1002/lpor.202400508}. Recent advances in fabrication processes and material platforms have substantially improved propagation losses, enabling the realization of increasingly complex photonic circuits in the visible spectral range for applications in fields such as quantum photonics~\cite{moody20222022}, sensing \cite{Stockinger:25,10.1117/1.APN.5.4.046001, Rank2021}, and nonlinear optics \cite{Zhao:20}. 

As the complexity of visible\-/light photonic circuits increases, the number of required components grows accordingly, placing greater demands on their performance and compactness. Nanophotonic inverse design offers a promising route to address these challenges by systematically optimizing device geometry and topology toward predefined target functions.
Through iterative optimization, inverse design enables the exploration of unconventional design spaces, leading to highly compact structures with improved performance beyond traditional approaches~\cite{Molesky2018, https://doi.org/10.1002/lpor.201000014}.

Although inverse design holds considerable potential for visible\-/light applications, its implementation in this spectral range remains at an early stage, with comparatively few demonstrations reported in the literature~\cite{Wang:26,Dory2019,Spektor:23,Zalogina2025}. This limited adoption can be attributed to several challenges inherent to visible\-/wavelength operation and the associated design and fabrication constraints. In particular, visible\-/light photonic platforms such as silicon nitride ($\mathrm{Si_3N_4}$) typically provide lower refractive index contrast compared to established silicon platforms, thereby constraining modal confinement and design flexibility~\cite{https://doi.org/10.1002/lpor.202400508, bi2026inverse, pita2025integrated, han2025breaking}. Furthermore, the shorter wavelengths involved require finer spatial discretization in electromagnetic simulations, resulting in substantially increased computational complexity and memory requirements, which pose significant challenges for nanophotonic device optimization~\cite{https://doi.org/10.1515/nanoph-2024-0127}.

In this work, we present an adjoint-based inverse design workflow for integrated photonic components tailored to visible\-/light applications. Specifically, we extend \textsc{fdtdx}~\cite{Mahlau2026}, an open-source Python finite-difference time-domain (FDTD) solver, with a fabrication\-/aware formalism~\cite{hiesener2025seeded} to optimize photonic components operating in the visible spectral range. \textsc{fdtdx} is particularly well-suited for optimizing visible\-/light photonic components with their associated challenges in computational complexity, as it leverages GPU acceleration for fast, memory\-/efficient simulations. 
To demonstrate the capabilities of our approach, we design, fabricate, and experimentally validate a set of photonic devices on a silicon nitride platform, optimized for broadband operation across the visible spectral range. A number of representative examples are investigated for routing and guiding, multiplexing, and polarization control.

The outline of the paper is as follows. In Sec.~\ref{sec:results}, we start by presenting the theory and implementation of our optimization workflow. We continue with the resulting inverse-designed photonic structures and discuss in detail their function and performance. In Sec.~\ref{sec:conclusion}, we summarize our findings and look ahead to future work. Finally, in Sec.~\ref{sec:methods} we describe the fabrication and experimental evaluation of photonic components. Some details underlying our optimization scheme and experimental evaluation are provided in the Supporting Information. 

\section{\label{sec:results}Results}
\begin{figure*}[t]
	\centering
	\includegraphics[width = 1\textwidth]{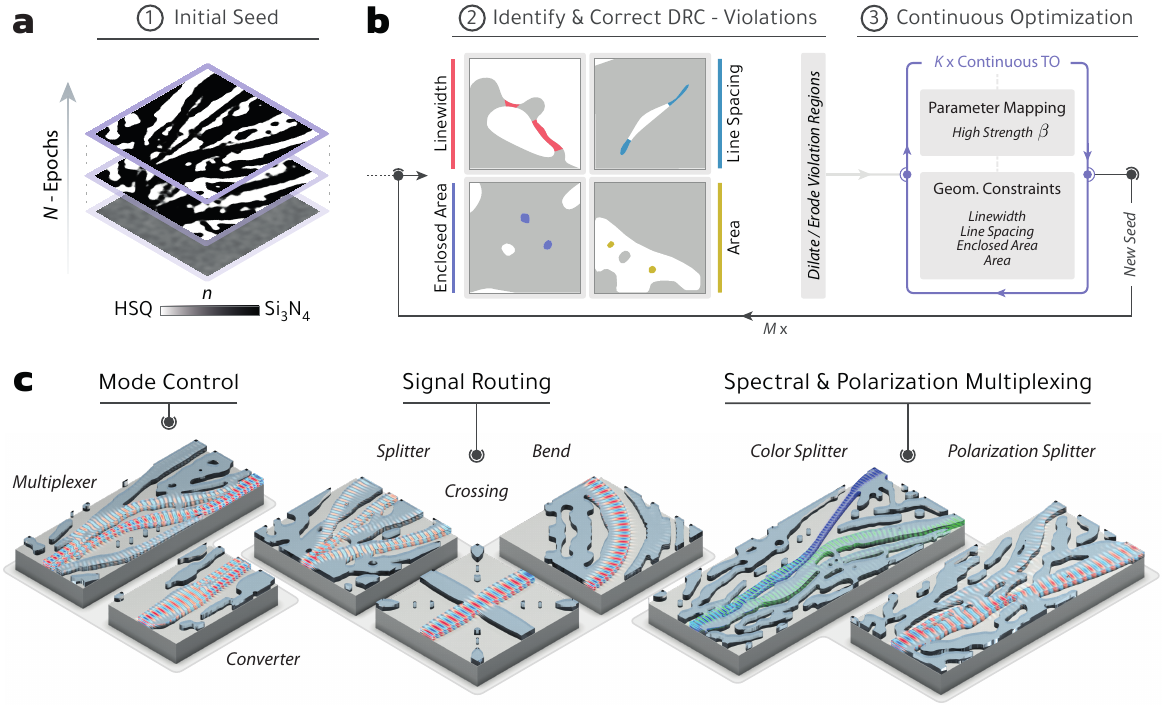}
	\caption{Schematics of seeded topology optimization for the fabrication-aware inverse design of integrated photonic devices. (a) In the first stage of optimization, an initial \textit{Seed} is generated through continuous topology optimization (TO; see also Fig.~\ref{fig:schematicsadjoint}). (b) The initial seed is inspected for design-rule check (DRC) violations, including minimum linewidth, line spacing, area, and enclosed area. Violating areas are corrected by adding or removing material, and the resulting material distribution is further optimized through TO under geometric line and area constraints and a high degree of binarization. The refined design then serves as the new seed for the subsequent DRC-correction and refinement cycle, which is repeated until a design-rule violation-free device is obtained or a maximum number of $M$ seeded cycles is reached. (c) Overview of the family of inverse-designed photonic devices for the on-chip processing and control of light modes.}
	\label{fig:schematicsseededto}
\end{figure*}
\subsection{Theory}
In this section, we start by discussing the electromagnetic aspects of the inverse-design framework used in this work. The framework is based on continuous topology optimization (TO) and incorporates iterative design-rule checks and corrections through the recently proposed seeded topology optimization~\cite{hiesener2025seeded}. Our aim is to convey the overall design workflow, rather than to discuss each numerical step in full detail. We first describe the electromagnetic formulation and modal analysis, followed by the continuous and seeded TO framework.

Common to both optimization schemes is the principle of adjoint-based optimization, which is summarized in Fig.~\ref{fig:schematicsadjoint} using an integrated power splitter as an example. The approach requires two electromagnetic simulations: a forward and an adjoint simulation. In the forward simulation (panel (a)), a supported eigenmode with frequency $\omega$ is launched into the device geometry through the input waveguide and propagated through the structure. The resulting electromagnetic field evolution is governed by the time-dependent Maxwell's equations,    
\begin{equation}
	\begin{gathered}
		\varepsilon(\bm r) \frac{\partial \bm E}{\partial t} = \nabla \times \bm H - \bm J \,, \\
		\mu_0 \frac{\partial \bm H}{\partial t} = - \nabla \times \bm E \,, \\
	\end{gathered}
\end{equation}
where $\varepsilon(\bm r)$ and $\mu_0$ are the spatial permittivity distribution and vacuum permeability, respectively, and $\bm J$ is the current density. To analyze the modal field distribution at a given frequency, we transform the time-dependent fields into their frequency-domain representation,
\begin{equation}\label{eq:fdfield}
    \tilde{\bm E}(\bm r, \omega) = \int_{t_0}^{t_1} \bm E(\bm r, t) e^{i \omega t} dt \,,
\end{equation}
where $\tilde{\bm{E}}$ denotes the complex electric-field phasor, with a similar expression for the magnetic field. Equation~\eqref{eq:fdfield} gives an estimate of the complex amplitude and phase of the electromagnetic fields, as obtained by integrating over a finite time window $t_1 - t_0$. Similarly, by projecting the frequency-domain field $\bm v$ onto a target mode $\bm u$ at the output waveguides, we obtain the complex mode overlap through the overlap integral,
\begin{equation}\label{eq:overlap}
    \mathcal{O}(\bm{u}, \bm{v}) = \frac{1}{4} \int_{A} \left(\tilde{\bm{E}}_1^* \times \tilde{\bm{H}}_2 + \tilde{\bm{E}}_2 \times \tilde{\bm{H}}_1^* \right) \cdot d\bm{A} \,,
\end{equation}
where $A$ denotes the monitor cross section with the infinitesimal area element $d\bm{A} = \hat{\bm{n}} \, dA$. Here the indices $1$ and $2$ correspond to the electric and magnetic field phasors of $\bm{u}$ and $\bm{v}$, respectively. The overlap $\mathcal{O}$ provides a measure of how well the propagated fields match a target mode in its spatial field distribution and complex amplitude. Finally, the corresponding power coupled into the target mode is obtained from the squared magnitude of the overlap,
\begin{equation}\label{eq:overlapnorm}
    \mathcal{P}_m = \left|\mathcal{O}(\bm{u}
    _m,\bm{v}_s)\right|^2 \,,
\end{equation}
with the source and target eigenmodes normalized to unit power. This expression serves as a quantitative performance metric, which can be compared with a desired output distribution to define a figure of merit (FOM), thereby completing the forward simulation. 

In the adjoint simulation, target modes are launched in reverse from the corresponding output ports and weighted according to the FOM, see Fig.~\ref{fig:schematicsadjoint}~(b). The resulting fields form the adjoint field $\bm{E}_{\mathrm{adj}}$. Its spatial overlap with the forward field in the design region yields the gradient of the FOM with respect to the material distribution,  
\begin{equation}
    \nabla_\varepsilon \mathrm{FOM} \propto \operatorname{Re}\left(\bm{E}_{fwd} \cdot \bm{E}_{adj}\right) \,.
\end{equation}
This gradient marks the central quantity used for the gradient-based optimization in this work and can be evaluated once the forward and adjoint fields are at hand. Its magnitude indicates how strongly the FOM responds to local permittivity changes (see Fig.~\ref{fig:schematicsadjoint}~(c)), while the sign determines whether increasing or decreasing the permittivity, i.e., adding or removing material, improves the FOM.
\subsection{Seeded Topology Optimization}

\begin{table*}[t]
	\caption{List of optimization and design parameters for inverse-designed structures. We indicate the target wavelengths $\mathrm{R}=638$~\si{\nano\meter}, $\mathrm{G}=523$~\si{\nano\meter}, $\mathrm{B}=458$~\si{\nano\meter}, and the number of conventional and seeded optimization epochs, respectively.}
	\label{tab:optparam}
	\begin{tabularx}{0.75\textwidth}{l r r r >{\raggedleft\arraybackslash}X}
		\hline\hline
		Device & RGB & Epochs Conv. TO ($N$) & Epochs Seeded TO ($M \times K$) & Size ($\mu\mathrm{m}^2$) \\
		\hline
        Bend & RGB & 80 & 8 x 15 & 4.5 x 4.5 \\
		Crossing & RGB & 100 & N/A & 4 x 4  \\
		Power Splitter & RGB & 100 & 8 x 15 & 5.5 x 5.5 \\
		Mode Converter & GB & 100 & 8 x 15 & 4.5 x 3 \\
		MDM & B & 160 & 8 x 15 & 8 x 3\\
        WDM & GB & 125 & 8 x 15 & 10 x 4\\
		Polarization Splitter & G & 100 & 8 x 15 & 9 x 4\\
		\hline
		\hline
	\end{tabularx}
\end{table*}

Building on the adjoint formulation from the previous section, we now describe how the resulting gradient is used in continuous topology optimization to iteratively update the design, as schematically shown in Fig.~\ref{fig:schematicsadjoint}~(c). In particular, we formulate our optimization problem as
\begin{equation}
    \begin{gathered}
        \max_\rho~\mathrm{FOM}\left(\rho(\bm r), \left\{\bm E_n(\bm r, t), \bm H_n(\bm r, t)\right\}_{t \in \left[t_0, t_1\right]} \right)\,, \\
        0 \leq \rho(\bm r) \leq 1, \quad n = 1, \ldots, N_\lambda \,,
    \end{gathered}
\end{equation}
where we have introduced $\rho(\bm r)$ as the continuous optimization variable in the design region, while $n$ indexes $N_\lambda$ wavelengths considered simultaneously during the optimization. 

Figure~\ref{fig:schematicsseededto}~(a) shows, as an example, the evolution of the inverse-designed power splitter starting from a random material distribution. At each optimization step, the design variable $\rho$ is updated to improve the target performance. This is supported by introducing material regularization steps~\cite{Hammond:21}, where we first apply a conic smoothing kernel $w$ to the optimization variable, such that
\begin{equation}
	\tilde{\rho} = w \ast \rho \,.
\end{equation}
In a second step, the filtered variable is projected toward a binary material distribution according to
\begin{equation}
	\bar{\rho} = \frac{\tanh\left(\beta \eta\right) + \tanh{\left(\beta \left(\tilde{\rho} - \eta\right)\right)}}{\tanh\left(\beta \eta\right) + \tanh\left(\beta (1 - \eta)\right)} \,,
\end{equation}
where $\beta$ and $\eta$ control the strength and threshold of the projection, respectively. Finally, $\bar{\rho}$ is mapped linearly onto the spatial permittivity distribution $\varepsilon(\bm r)$ between the core and cladding materials. The two-dimensional material distribution is extruded uniformly along the vertical dimension ($+\hat{z}$), resulting in a constant device height throughout the design region. Over $N$ optimization epochs, $\beta$ is continuously increased to promote a nearly binary structure. This, in turn, yields an \textit{Initial Seed} that serves as a starting point for seeded TO.

While the initial seed in our example features a high degree of binarization, it may still contain features that violate prescribed fabrication rules. Therefore, the initial seed is subjected to a design-rule check (DRC) following the seeded TO procedure introduced in Ref.~\cite{hiesener2025seeded}, as illustrated in panel (b) of Fig.~\ref{fig:schematicsseededto}. We identify regions violating minimum-feature or geometric constraints, i.e., minimum linewidth, line spacing, enclosed area, and area, using indicator functions. As a second step, violating regions are corrected locally by eroding the core material to comply with minimum-line spacing and enclosed-area specifications. Likewise, minimum-linewidth and minimum-area violations are corrected by dilating the core material, reducing the number of DRC violations. For a detailed description of the correction procedure based on indicator functions, we refer the reader to Refs.~\cite{hiesener2025seeded, Hammond:21}.

Because this correction step is performed regardless of the FOM, the optical performance of the corrected structure may be perturbed. Therefore, the DRC-corrected design is subsequently used as the starting point for a new continuous TO stage, with the corrected structure reinitialized as the optimization variable; see also Fig.~\ref{fig:schematicsseededto}~(b). Over $K$ iterations, the performance of the design is further improved with respect to the FOM while enforcing strong binarization and geometric constraints, with the latter helping to preserve the corrected topology. We repeat this optimization process until a design-rule violation-free device is obtained or a maximum number of cycles $M$ is reached. 
\begin{figure*}[t]
    \centering
    \includegraphics[width = 1\textwidth]{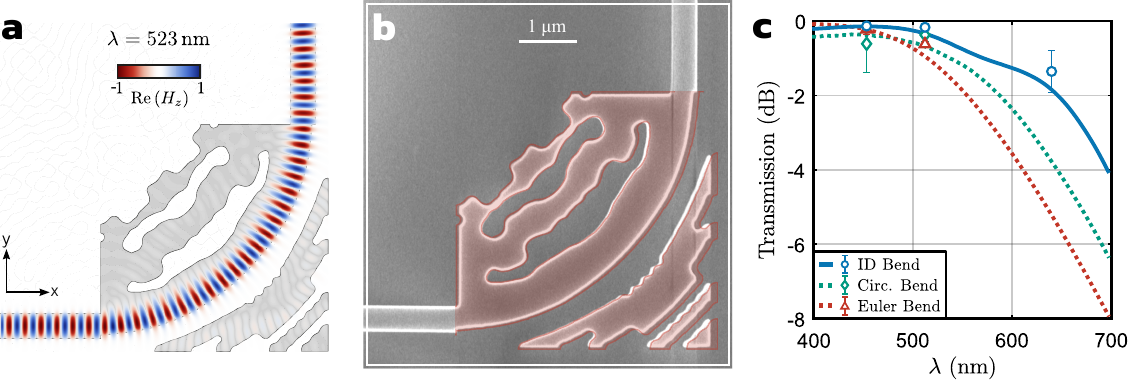}
\caption[]{Inverse\-/designed waveguide bend. (a) Final design of the optimized waveguide bend with an effective radius of \SI{4}{\micro\meter}, along with the superimposed field distribution ($H_\mathrm{z}$) for the fundamental TE\-/mode at a wavelength of $\lambda = 523$~\si{\nano\meter}. The gray regions represent the waveguide core material ($\mathrm{Si_3N_4}$), whereas the white regions indicate the cladding material (HSQ). (b) SEM image of the fabricated waveguide bend, overlaid with a semi-transparent mask of the inverse-designed component for comparison. (c) Simulated and measured transmission of the inversely designed bend (blue) over the wavelength range from 400 to \SI{700}{\nano\meter}. The performance of the optimized device is benchmarked against an Euler bend (red) and a circular bend (green), both featuring a radius of \SI{4.5}{\micro\meter}.}
    \label{fig: waveguide bend}
\end{figure*}
\subsection{\label{sec:}Inverse Design of Photonic Components}
Using the methodology discussed in the previous section, we design and simulate topology\-/optimized integrated photonic components with the open\-/source Python finite-difference time-domain solver \textsc{fdtdx}. The design optimization is performed using the gradient\-/based \textsc{nadam} method \cite{Dozat2016}, as implemented in the Python library \textsc{optax} \cite{Bradbury2018JAX}. In this process, gradient\-/based optimization is employed to maximize a figure of merit ($\mathrm{FOM}$), typically defined by the transmission into specific output ports of the device. A monochromatic continuous wave source is used to inject the desired waveguide mode into the component's input waveguide. The transmission into the target output port is quantified via the spatial overlap of the simulated field at the output monitor with the corresponding waveguide eigenmode field, normalized by the source power (see Eqs.~\eqref{eq:overlap} and \eqref{eq:overlapnorm}). To avoid parasitic effects due to source ramping, the simulation is run until the fields are fully developed, with at least ten additional periods used for measurements. The detector fields are integrated over the final ten periods, see Eq.~\eqref{eq:fdfield}. Furthermore, to optimize component performance over a broad spectral range, simulations are performed at multiple discrete wavelengths during each iteration of the optimization process, specifically $\lambda_R = 638$~\si{\nano\meter}, $\lambda_G = 523$~\si{\nano\meter}, and $\lambda_B = 458$~\si{\nano\meter}. All optimizations are carried out using a uniform mesh with a resolution of 20~\si{\nano\meter}, providing sufficient accuracy in the spectral range under investigation.
 
Constant refractive indices are used across all wavelengths: ${n_\mathrm{SiO_2} = 1.46}$ for the silica substrate, ${n_\mathrm{Si_3N_4} = 2.03}$ for the silicon nitride waveguide structures, and ${n_\mathrm{HSQ} = 1.45}$ for the HSQ cladding. Although neglecting dispersion may reduce simulation accuracy, the selected materials exhibit sufficiently weak dispersion over the wavelength range considered. To justify this assumption, we have performed verification simulations, including dispersion, using Ansys' Lumerical and Flexcompute's Tidy3D, both exhibiting excellent agreement with \textsc{fdtdx} for all components. 

In this work, we applied our design methodology to a variety of components to highlight its wide applicability. Figure~\ref{fig:schematicsseededto}~(c) provides an overview of the optimized components, and Table~\ref{tab:optparam} lists the main design parameters, including operating wavelengths, device dimensions, and the number of iterations used for the continuous and seeded TO. All photonic devices were optimized enforcing a minimum feature size of \SI{100}{\nano\meter} to ensure manufacturability.

All components were fabricated in silicon nitride with a constant layer thickness of \SI{100}{\nano\meter} and connecting waveguides feature a constant width of \SI{500}{\nano\meter}. The fabricated components were inspected by scanning electron microscopy (SEM) and the resulting micrographs are presented below with overlays of the corresponding inverse-designed geometries (transparent red) to assess fabrication fidelity. Although imaging artifacts introduce minor distortions, the micrographs provide a qualitative assessment of the fabricated structures.
Furthermore, the performance of the devices was evaluated both experimentally and numerically. Cascaded configurations of the respective devices were characterized experimentally to determine their insertion losses and cross-talk behavior. For devices with more complex functionalities, additional measurement configurations were employed, as detailed in the Supporting Information, Secs. 1A–G. The experimental results and corresponding simulations performed using Tidy3D are presented below for each respective device.

In the following, we discuss each component in detail. We begin with components that primarily enable targeted light routing, such as optimized waveguide bends, crossings, and a power splitter with a custom splitting ratio. Next, we consider components that specifically manipulate the spatial waveguide mode, focusing on a mode converter and a mode division multiplexer. Finally, we investigate components enabling selective control over wavelength and polarization in waveguides, such as wavelength division multiplexers and polarization splitters.

\begin{figure*}[t]
    \centering    
    \includegraphics[width=1\textwidth]{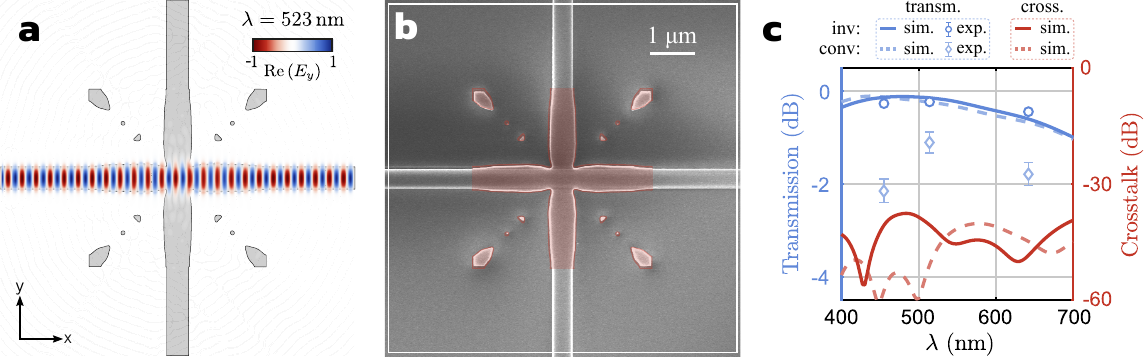}
\caption[]{Inverse-designed waveguide crossing. (a) Optimized waveguide crossing design with a $4 \times 4$~\si{\micro\meter\squared} footprint, showing the electric field distribution ($E_\mathrm{y}$) of the fundamental TE mode at a wavelength of $\lambda = 523$~\si{\nano\meter}. (b) SEM image of the fabricated waveguide crossing. (c) Simulated and measured transmission (blue) and crosstalk (red) of the waveguide crossing over the wavelength range 400–700~\si{\nano\meter}. The performance of the optimized device is benchmarked against a conventionally designed crossing optimized via a particle swarm algorithm.}
    \label{fig: crossing}
\end{figure*}

\subsubsection*{Waveguide Bend}
Redirecting light efficiently is a fundamental requirement in many photonic circuits, making waveguide bends essential components in integrated photonics. In silicon nitride platforms, moderate refractive index contrast leads to the use of large bend radii to minimize radiation losses, ultimately restricting integration density~\cite{Vogelbacher:19,Zhang:23}. Here, we present an inverse-designed silicon nitride waveguide bend with a compact effective radius of 4~\si{\micro\meter}, optimized to guide the fundamental TE mode across the entire visible spectrum. The design region for the bend was specified as $4.5 \times 4.5$~\si{\micro\meter\squared}.

A schematic representation of the final design, along with the superimposed field distribution ($\lambda = 523$~\si{\nano\meter}), is shown in Fig.~\ref{fig: waveguide bend}~(a). The corresponding scanning electron microscope image of the fabricated device is depicted in panel (b) of Fig.~\ref{fig: waveguide bend}. Upon detailed inspection of the design, several key features emerge that contribute to an intuitive understanding of the mechanisms behind loss reduction. First, a lateral offset between the bend and the input and output waveguides is observed. This feature has been reported to reduce losses associated with mode mismatches at the interface between straight and curved waveguide sections~\cite{251169,372446}. Furthermore, Bragg\-/mirror\-/like structures are evident along the outer edge of the bend, which serve to suppress radiation losses.

The performance of the optimized waveguide bend design was evaluated by comparing it to conventional bend geometries. Specifically, circular and Euler bends with a radius of \SI{4.5}{\micro\meter} were considered. To enable a comprehensive evaluation, both numerical simulations and experimental measurements were performed. 
The transmission spectra of the different bend designs over a wavelength range from 400 to \SI{700}{\nano\meter} are shown in Fig.~\ref{fig: waveguide bend} (c). Experimental results are shown as markers, while the corresponding simulation results are shown as solid and dashed curves. All bend designs exhibit low insertion losses at shorter wavelengths, however, the losses increase substantially toward longer wavelengths. For conventional bend designs, the resulting high insertion losses at the red wavelength led to transmitted power levels below the experimental noise floor, preventing a reliable quantitative determination of the insertion loss. Consequently, no experimental data are shown for the conventional designs at this wavelength. 
Notably, the topology optimized design outperformed conventional designs across all three target wavelengths, achieving experimentally measured losses of -$0.08\pm0.04$~\si{\decibel}, -$0.11\pm0.08$~\si{\decibel} and -$1.3\pm0.6$~\si{\decibel} at 455, 514, and 642~\si{\nano\meter}, respectively.
The device transmission was extracted from a fit to the experimentally measured transmission of cascaded structures with varying numbers of devices. The associated uncertainties represent the 95\% confidence intervals of the fitted transmission coefficients.

\subsubsection*{Waveguide Crossing}

Compact and low-loss waveguide crossings are fundamental building blocks for dense and scalable photonic circuits. 
Conventional strategies to reduce losses in waveguide crossings typically employ multimode interference regions~\cite{Kim:14,Johnson:20} and inverted tapers~\cite{Han:22} within the crossing area. Although these approaches deliver promising results, they often depend on parameter sweeps, which can be computationally intensive. This work investigates a compact, topology-optimized silicon nitride waveguide crossing for low-loss transmission of the fundamental TE mode across the visible spectrum. The design region is restricted to a compact footprint of $4 \times 4$~\si{\micro\meter\squared},  exploiting the eightfold mirror symmetry of a square crossing design.
Feature-size constraints and seeded topology-optimization steps are not applied to this component, as prior experience indicates that small functional features do not emerge in the design, thereby accelerating the optimization process.

A schematic of the optimized design is shown in Fig.~\ref{fig: crossing} (a). The design incorporates a modulation of the waveguide geometry, in which the width is increased and then reduced immediately before the intersection. This geometric modulation concentrates the guided mode at the crossing, reducing mode spreading and suppressing crosstalk, similar to the approach described in Ref.~\cite{Han:22}.
\begin{figure*}[t]
    \centering
    \includegraphics[width = 1\textwidth]{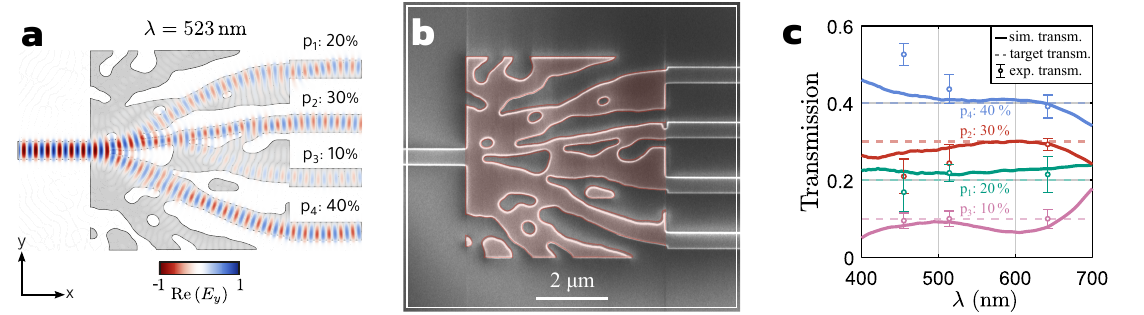}
    \caption[]{Inverse\-/designed 1 x 4 power splitter with specified power splitting ratio. (a) Optimized power splitter design with a $5.5 \times 5.5$~\si{\micro\meter\squared} footprint, showing the electric field distribution ($E_y$) at a wavelength of $\lambda$ = 523~\si{\nano\meter}. (b) SEM image of the fabricated power splitter, overlaid with a semitransparent representation of the design for direct comparison. (c) Simulated and measured transmission of the individual ports, normalized to the total transmission of the power splitter, over the wavelength range 400–700~\si{\nano\meter}. The experimentally measured total transmission of the powersplitter is -$3.62$~\si{\decibel}, -$2.06$~\si{\decibel} and -$1.37$~\si{\decibel} at 455, 514 and 642~\si{\nano\meter}, respectively.}
    \label{fig: power splitter}
\end{figure*}
\begin{figure*}[t]
    \centering
    \includegraphics[width = 1\textwidth]{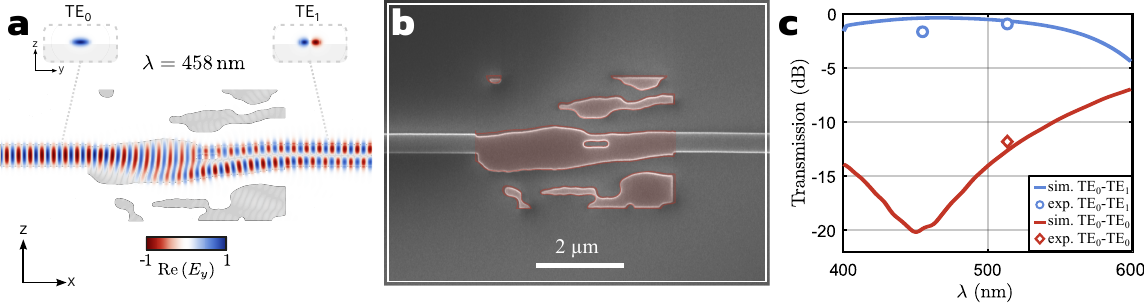}
    \caption[]{Inverse\-/designed mode converter. (a) Final design of the optimized $\mathrm{TE_0\,-\,TE_1}$ mode converter, along with the superimposed field distribution ($E_y$) at a wavelength of $\lambda = 458$~\si{\nano\meter}. (b) SEM image of the fabricated mode converter. (c) Simulated and measured transmission spectra of the $\mathrm{TE_0\,-\,TE_1}$ mode converter. The blue curve corresponds to the conversion of the $\mathrm{TE_0}$ mode to the $\mathrm{TE_1}$ mode, and the red curve represents the residual intensity in the $\mathrm{TE_0}$ mode.}
    \label{fig: mode converter}
\end{figure*}
The performance of the inverse\-/designed crossing is compared, both in simulation and experiment, with a conventional crossing design. The conventional crossing design employs a taper region followed by a multimode waveguide region~\cite{Sacher:19}. A particle swarm optimization algorithm was employed to optimize the device performance, with a footprint of $4 \times 4$~\si{\micro\meter\squared}. The optimization was performed over five iterations with a swarm size of 200 particles, resulting in a total of 1000 device simulations using Lumerical's \textsc{varFDTD} solver.

The transmission and crosstalk of the two crossing designs are shown in Fig.~\ref{fig: crossing} (c). 
Experimental results are shown as markers, while the corresponding simulation results are represented by solid and dashed curves. For all investigated devices, however, the crosstalk signal remained below the experimental noise floor, preventing reliable quantitative evaluation. Consequently, only the simulated crosstalk is shown in Fig.~\ref{fig: crossing} (c).
While the two designs exhibit comparable performance in simulation, the inverse\-/designed crossing demonstrates superior experimental performance across the entire design wavelength range, with insertion losses of -0.13~\si{\decibel}, -0.16~\si{\decibel}, and -0.53~\si{\decibel} at wavelengths of 455, 514, and 642~\si{\nano\meter}, respectively. Furthermore, simulation results indicate that both designs maintain crosstalk levels below \SI{-40}{\decibel} throughout the design wavelength range.

The degraded experimental performance of the conventional design suggests reduced robustness against fabrication-induced deviations, such as feature shrinkage, which was observed in the fabricated devices investigated in this work.
Furthermore, when comparing the two designs, we emphasize that the inverse-designed topology was obtained without any a priori information about the target geometry. Moreover, the gradient-based optimization employed for the inverse design offers a convergence substantially faster than that for particle swarm optimization.

\subsubsection*{Power Splitter}
Beyond efficient routing of guided modes, controlled power splitting among multiple channels is a central functionality in integrated photonics. Power splitters with tailored splitting ratios are, for example, employed in nanophotonic phased arrays~\cite{Sun2013,Poulton:17} and for on-chip power monitoring~\cite{481118}.
Using inverse design methodologies, power splitters with arbitrary splitting ratios can be precisely engineered to operate across multiple target wavelengths. This can be achieved by optimizing a suitable figure of merit, with one potential choice being:
\begin{equation}
\begin{split}
\text{FOM} =
\sum_{\lambda} \left[
\mathcal{P}^{(\lambda)}
- \sum_{j}
\left| \mathcal{P}_j^{(\lambda)} - p_j \, \mathcal{P}^{(\lambda)} \right|
\right],
\end{split}
\end{equation}
\noindent
where $\lambda$ denotes the set of wavelengths under consideration, $j$ is the set of output ports, $\mathcal{P}_j^{(\lambda)}$ represents the transmission measured in the fundamental TE mode at port $j$ for wavelength $\lambda$. Here $p_j$ is the desired fraction of the total transmission assigned to port $j$, and $\mathcal{P}^{(\lambda)} = \sum_j \mathcal{P}_j^{(\lambda)}$ is the total transmission of the device for the given wavelength.

A further consideration regarding the figure of merit introduced above concerns the non\-/differentiability of the modulus term at zero. Although this may, in principle, affect convergence, no such effects were observed in the present optimization. Nevertheless, a smooth metric, such as the variance, could potentially provide improved optimization performance.

We present an inverse-designed 1 × 4 power splitter with a specified output power splitting ratio optimized for RGB wavelengths. The output ports of the splitter are configured to deliver 10\%, 20\%, 30\%, and 40\% of the transmitted power. A schematic of the optimized design, along with the superposed field distribution, is presented in Fig.~\ref{fig: power splitter} (a), while a scanning electron microscope image of the fabricated device is shown in Fig. \ref{fig: power splitter} (b). Closer inspection of the device features suggests an intuitive interpretation in which the structure operates similar to a cascade of asymmetric Y-splitters, which together produce the desired output intensity distribution.

Figure \ref{fig: power splitter}(c) presents the simulated and experimentally measured transmission of each port of the power splitter, with all output powers normalized to the total transmitted power over the wavelength range of 400–700~\si{\nano\meter}.
The target transmission for each port is indicated by a dashed line, while the simulated transmission is shown as a solid line and the experimental data as markers. The experimental splitting ratio was determined from five individual power-splitter devices. The markers represent the mean splitting ratio across these devices, with the error bars indicating the corresponding standard deviation.
It is evident that the splitting ratio remains largely wavelength-independent across the selected design wavelength range, with the experimental results showing good agreement with the corresponding simulations. Furthermore, total insertion losses of $-3.62$~\si{\decibel}, $-2.06$~\si{\decibel} and $-1.37$~\si{\decibel} were measured at wavelengths of 455, 514, and 642~\si{\nano\meter}, respectively.


\subsubsection*{Mode Converter}
Precise control over the transverse spatial modes in waveguides enables a wide range of applications in integrated photonics, including dense optical data transmission~\cite{Stern:15} and advanced wavefront shaping and steering~\cite{Harrington:25,Zhao2025_AllOnChip}. In this context, mode converters are essential components, as they enable coupling between distinct spatial modes.
As part of this work, we investigate a mode converter designed to transform the fundamental $\mathrm{TE_0}$ mode into the first order $\mathrm{TE_1}$ mode. The converter is optimized for operation at wavelengths of $\lambda_B = 458$~\si{\nano\meter} and $\lambda_G = 523$~\si{\nano\meter}. Within this spectral range, the selected waveguide platform enables low-loss propagation of both the $\mathrm{TE_0}$ and $\mathrm{TE_1}$ modes. Moreover, the device features a highly compact footprint, occupying a total area of only $4.5 \times 3$~\si{\micro\meter\squared}.

Figure~\ref{fig: mode converter} (a) illustrates a graphical representation of the final design of the mode converter with the real\-/valued field distributions of the $E_\mathrm{y}$ component superimposed. Figure~\ref{fig: mode converter} (b) shows the SEM image of the manufactured component. An intuitive understanding of the optimized design can be obtained by examining its structural features. An asymmetric waveguide taper at the input forms a widened central waveguide region and introduces an initial lateral phase asymmetry in the guided mode. Within this widened section, an elongated and laterally offset hole further enhances the relative phase accumulation across the waveguide. The combined effect of these features produces an overall lateral phase shift of $\pi$, thereby enabling an efficient conversion to the $\mathrm{TE_1}$ mode.

Figure~\ref{fig: mode converter} (c) shows the simulated and measured transmission of the fundamental and first higher-order modes of the mode converter for an input fundamental TE mode. Transmission values of $-1.64$~\si{\decibel} and $-0.91$~\si{\decibel} were measured in the first-order mode at 455 and 514~\si{\nano\meter}, respectively. The residual power in the fundamental TE$_0$ mode was measured as $-11.8$~\si{\decibel} at 514~\si{\nano\meter}. At 455~\si{\nano\meter}, the corresponding signal was below the experimental noise floor, preventing reliable quantitative measurement. The experimentally measured conversion and residual fundamental-mode powers correspond to a conversion efficiency of $81\%$ and a conversion purity of $92.5\%$ at a wavelength of $514$~\si{\nano\meter}.


\subsubsection*{Mode Division Multiplexer}
\begin{figure*}[t]
    \centering
    \includegraphics[width = 1\textwidth]{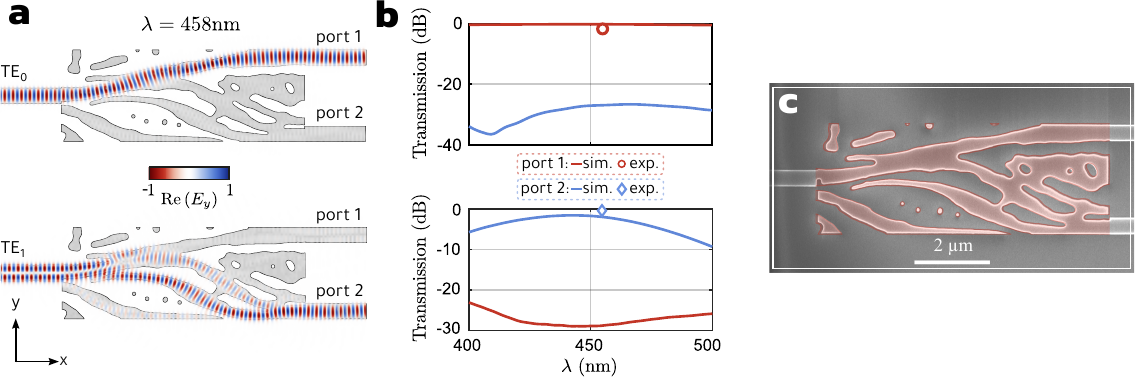}
\caption[]{Inverse\-/designed mode-division multiplexer (MDM). (a) Final design of the optimized $\mathrm{TE_0\,-\,TE_1}$ MDM, with the superimposed field distribution of the real part of the $\mathrm{E_y}$ componentent for input $\mathrm{TE_0\,-\,TE_1}$ modes. (b) Simulated and experimentally measured transmission spectra of the mode-division multiplexer for both input scenarios. The simulated transmission values were calculated by evaluating the mode overlap with the fundamental TE mode at the output ports. (c) SEM image of the fabricated device overlaid with a semitransparent representation of the design for direct comparison.}
    \label{fig: mode sorter}
\end{figure*}
Complementary to targeted mode conversion, we further investigate mode separation and multiplexing through a (de)multiplexer designed to isolate or combine the fundamental and first-order TE modes. The device converts both modes into fundamental TE modes and routes them to two distinct output ports, and vice versa.
The mode demultiplexer is optimized for operation at a wavelength of $\lambda_\mathrm{B} = 458$~\si{\nano\meter}. It features a compact device footprint of $8 \times 3$~\si{\micro\meter\squared}.
The figure of merit used in the optimization process is defined as a weighted combination of the mean transmitted power in the target port and penalties associated with port imbalance and crosstalk,
\begin{equation}
\begin{aligned}
\mathrm{FOM} ={}&
w_1\overline{\mathcal{P}}_{\mathrm{th}}
-w_2\operatorname{Var}\!\left(\mathcal{P}_{\mathrm{th}}\right)\\
&-w_3\overline{\mathcal{P}}_{\mathrm{x}}
-w_4\operatorname{Var}\!\left(\mathcal{P}_{\mathrm{x}}\right).
\end{aligned}
\end{equation}
Here, $\overline{\mathcal{P}}_\mathrm{th}$ denotes the mean transmitted power in the target (through) port, while $\mathrm{Var}\!\left(\mathcal{P}_{\mathrm{th}}\right)$ represents its variance over the considered design wavelengths and input mode scenarios. Similarly, $\overline{\mathcal{P}}_\mathrm{x}$ and $\mathrm{Var}\!\left(\mathcal{P}_{\mathrm{x}}\right)$ correspond to the mean and variance of the crosstalk power coupled to the undesired ports. 
The weight factors $w_1$-$w_4$ are real-valued coefficients that allow the optimization to balance the penalties associated with fluctuations and crosstalk against overall transmission performance.

A graphical representation of the final mode (de)multiplexer design, including overlaid field amplitudes of the $E_\mathrm{y}$ component for input TE$_0$ and TE$_1$ modes, as well as a scanning electron microscope (SEM) image of the fabricated device, is shown in Figure \ref{fig: mode sorter} (a) and (c).


Figure~\ref{fig: mode sorter}~(b) shows the experimentally measured and simulated transmission characteristics for excitation of TE$_0$ and TE$_1$ input modes. The simulation results are represented by solid curves, while the experimental data are shown as markers. For both input modes, the transmitted power at the through port was experimentally characterized, with a transmission of $-1.83$~\si{\decibel} measured at port 1 for TE$_0$ excitation and a transmission of $-0.45$~\si{\decibel} at port 2 for TE$_1$ excitation at a wavelength of 455~\si{\nano\meter}. In contrast, the transmission into the cross port could not be reliably quantified, as the corresponding signals remained below the experimental noise floor.
\subsubsection*{Wavelength Division Multiplexer}
\begin{figure*}[t]
    \centering
    \includegraphics[width = 1\textwidth]{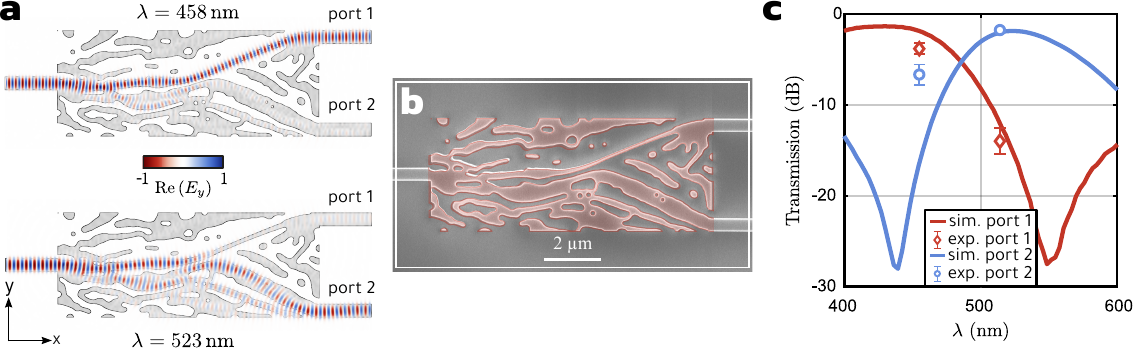}
\caption[]{Inverse-designed wavelength-division (de)multiplexer (WDM). (a) Final design of the optimized WDM, with the superimposed field distribution of the real part of the $\mathrm{E_y}$ component for input wavelengths of $\lambda_\mathrm{B} = 458$~\si{\nano\meter} and $\lambda_\mathrm{G} = 523$~\si{\nano\meter}. (b) SEM image of the fabricated WDM overlaid with a semitransparent representation of the design for direct comparison. (c) Simulated and measured transmission spectra of the WDM. The simulated transmission values were calculated by evaluating the mode overlap with the fundamental TE mode at the output ports. }
    \label{fig: color sorter}
\end{figure*}

Another property of light that we aim to control using inverse-designed integrated photonic structures is its wavelength. In particular, different wavelengths can be selectively routed into separate paths or combined into a common channel, which is highly relevant for applications such as on-chip spectrometry \cite{Li2022,Ryckeboer:13,Cheben:07,Xia:11} and high-density data transmission \cite{Gao:17,Sabri:24}. To demonstrate wavelength-selective routing, we design a wavelength-division (de)multiplexer that separates two target wavelengths, $\lambda_\mathrm{G} = 523$~\si{\nano\meter} and $\lambda_\mathrm{B} = 458$~\si{\nano\meter}, into distinct ports, using the fundamental TE mode at both the input and output ports. The design footprint is set to $10 \times 4$~\si{\micro\meter\squared}.
For the optimization process, we employed a figure of merit closely related to that used for the mode-division multiplexer, in which the average transmission to the target ports is maximized while port imbalance and crosstalk are minimized,
\begin{equation}
\begin{split}
\text{FOM} = w_1\overline{\mathcal{P}}_{\mathrm{th}}-w_2\mathrm{Var}\!\left(\mathcal{P}_{\mathrm{th}}\right)-w_3\overline{\mathcal{P}}_{\mathrm{x}}
\end{split}\,,
\end{equation}
where $\overline{\mathcal{P}}_\mathrm{th}$ denotes the mean transmitted power in the target (through) port, while $\mathrm{Var}\!\left(\mathcal{P}_{\mathrm{th}}\right)$ represents its variance over the considered design wavelengths. Similarly, $\overline{\mathcal{P}}_\mathrm{x}$ corresponds to the mean crosstalk power coupled to the undesired port.
Similarly, the weight factors $w_1$-$w_3$ are real-valued coefficients that allow for balanced penalties associated with fluctuations and crosstalk against overall transmission performance.

The optimized device topology, together with the superposed field distributions of the real part of the E$_\mathrm{y}$ component for input light at $\lambda_B = 458$~\si{\nano\meter} and $\lambda_G = 523$~\si{\nano\meter}, is depicted in Fig. \ref{fig: color sorter} (a). Figure \ref{fig: color sorter} (b) shows an SEM image of the fabricated device with a semitransparent overlay of the design topology for comparison.

The experimentally measured and simulated transmission characteristics of the wavelength division (de)multiplexer over a wavelength range from 400 to 600~\si{\nano\meter} are shown in Fig.~\ref{fig: color sorter}(c). The simulation results are represented by solid curves, while the experimental data are shown as markers. The overall transmission was characterized using cascaded devices, while the relative transmission at each output port was determined from measurements of five individual devices. The markers represent the absolute transmission at each output port, calculated by combining the mean relative transmission obtained from five individual devices with the corresponding total transmission measured from the cascaded devices. The error bars indicate the standard deviation of the relative transmission measured across the five individual devices.

While the measurements and simulations show close agreement at 514~\si{\nano\meter}, deviations become apparent at the shorter wavelength of 455~\si{\nano\meter}. In particular, both the overall transmission and the sorting fidelity are reduced compared with the simulated results. Although several factors may contribute to these deviations, feature shrinkage observed in the fabricated structures represents one possible source of the discrepancy.

\subsubsection*{Polarization Splitter}
\begin{figure*}[t]
    \centering
    \includegraphics[width = 1\textwidth]{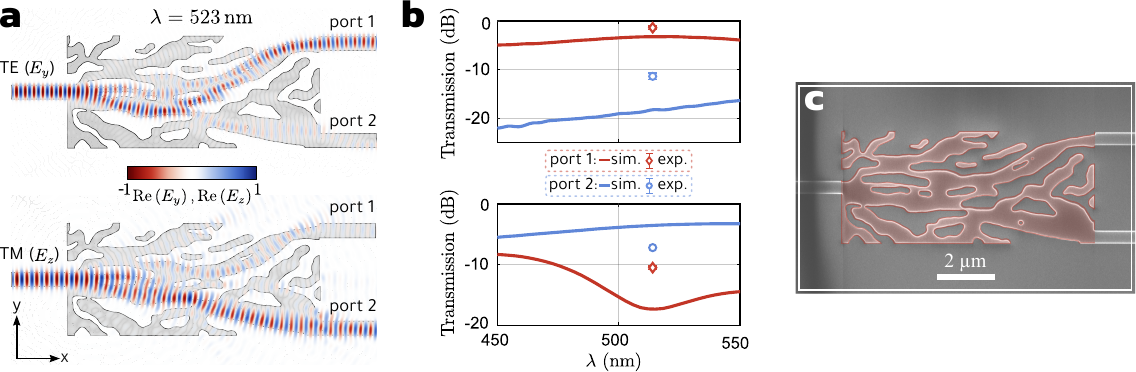}
\caption[]{Inverse-designed polarization splitter. (a) Final design of the optimized polarization splitter, together with superposed field distributions of the real value of the $E_y$ component for the input cases of fundamental TE and TM modes. (b) Simulated and measured transmission spectra of the polarization splitter for the cases of fundamental TE and TM modes as input. The simulated transmission values were calculated by evaluating the mode overlap with the fundamental TE and TM modes at the output ports for the respective input cases. (c) SEM image of the fabricated polarization splitter overlaid with a semitransparent representation of the design for direct comparison.}
    \label{fig: pol. sorter}
\end{figure*}
Finally, we consider a component designed to tailor the polarization properties of light in waveguide modes. Specifically, we analyze the splitting or combination of orthogonal polarization states, analogous to a polarization beam splitter in conventional optics. These components enable wide range of applications including quantum key distribution \cite{Semenenko:20,Minder2019}, polarization-encoded communication \cite{8467540}, and integrated polarimetry \cite{Zhao:24,Fang:21}. Here, we consider a polarization splitter designed to separate the fundamental TE and TM modes and route them to distinct output ports. Operation is optimized for a design wavelength of $\lambda_\mathrm{G} = 523$~\si{\nano\meter} within a compact $9 \times 4$~\si{\micro\meter\squared} footprint. 
A simple figure of merit was defined to maximize balanced transmission to the desired ports while minimizing crosstalk to the undesired ports with the expression,
\begin{equation}
\begin{split}
\text{FOM} = \overline{\mathcal{P}}_{\mathrm{th}} - \overline{\mathcal{P}}_{\mathrm{x}} - \operatorname{std}\left(\mathcal{P}_{\mathrm{th}}\right) - \operatorname{std}\left(\mathcal{P}_{\mathrm{x}}\right)
\end{split} \,,
\end{equation}
where $\overline{\mathcal{P}}_\mathrm{th}$ denotes the mean transmitted power in the target port (through), while $\operatorname{std}\!\left(\mathcal{P}_{\mathrm{th}}\right)$ represents its standard deviation over the two polarization cases considered. Similarly, $\overline{\mathcal{P}}_\mathrm{x}$ corresponds to the mean crosstalk power coupled to the undesired port and $\operatorname{std}\!\left(\mathcal{P}_{\mathrm{x}}\right)$ its standard deviation across the two polarization states.

Figure \ref{fig: pol. sorter} (a) shows the optimized device geometry with superimposed field distributions of the real part of the $E_y$ component for TE and $E_z$ for TM polarized fundamental modes, respectively, at a wavelength of $\lambda_\mathrm{G} = 523$~\si{\nano\meter} under different input polarization conditions. The fabricated device was imaged using an SEM, and a semi-transparent overlay illustrates its desired geometry (Figure~\ref{fig: pol. sorter}~(c)).

The experimentally measured and simulated transmission characteristics of the polarization splitter are shown in Fig.~\ref{fig: pol. sorter}~(b). The simulation results are represented by solid curves, while the experimental data are shown as markers. The overall transmission was determined by fitting measurements obtained from cascaded structures containing different numbers of devices. The relative transmission between the two output ports was characterized using five individual devices. The markers represent the resulting combination of the fitted overall transmission and the mean relative transmission obtained from the five devices, while the error bars indicate the standard deviation of the relative transmission across these devices.

For TE\-/polarized excitation, the experimentally measured transmission into port 1 shows good agreement with the corresponding simulation results, whereas the experimentally measured crosstalk into port 2 is higher than predicted. For TM\-/polarized excitation, both the transmission into port 2 and the crosstalk into port 1 exhibit deviations from the simulated values, indicating reduced performance. As observed for the other components, feature shrinkage was present in the fabricated polarization splitter. This fabrication\-/induced deviation from the nominal geometry may contribute to the observed discrepancies between experiment and simulation.  

\section{\label{sec:conclusion}Discussion}
Silicon nitride integrated photonic components for the visible spectrum were inverse-designed and validated using FDTD simulations and experiments, with a subset of components benchmarked against conventional design approaches.
The designed components cover a broad range of functionalities for the control and processing of on-chip light modes. Devices focused on low-loss routing of waveguide modes include waveguide bends, waveguide crossings, and variable power splitters. Beyond low-loss routing, the design approach was applied to components targeting spatial modes in waveguides, including mode converters and mode division multiplexers. Furthermore, a wavelength-division multiplexer and a polarization splitter were designed and investigated.
The design process leveraged the open-source FDTD solver \textsc{fdtdx}, which was extended to support a seeded topology optimization procedure in order to satisfy common design-rule constraints and ensure robust manufacturability using standard fabrication methods. Moreover, broadband performance across the visible spectrum was achieved through parallel optimization at discrete wavelengths.

Our findings demonstrate that the presented inverse design approach enables targeted control of light parameters in silicon nitride waveguides with high efficiencies throughout the visible spectrum. Notably, the approach can be applied to an extended range of integrated photonic components and applications across the visible to infrared spectrum, including grating couplers~\cite{Efseaff:24, Hammond:22,Huang2025} and photonic computing~\cite{Sved2026, Zhao2025,Nikkhah2024}. 
In the future we plan to extend our design method to support topology\-/optimized multilayer systems~\cite{Hammond:22}, potentially enabling additional functionality and improved device performance. Moreover, fabrication variations such as over- or under-etching will be incorporated into the optimization process to produce designs with even more enhanced robustness against fabrication tolerances~\cite{Hammond:21}. Future implementations may further include dispersion-aware optimization, for precise tuning of broadband performance and greater control over chromatic behavior.

The inverse\-/designed components developed in this work, along with the associated design process, demonstrate a route to compact and high\-/performance integration in silicon nitride\-/based photonic systems for visible\-/light applications.

\section{Methods}\label{sec:methods}

\subsection{Fabrication}
The photonic integrated chip was fabricated on a material platform purchased from \textit{LioniX international}, consisting of a 100-\si{\nano\meter}-thick $\mathrm{Si_3N_4}$ film deposited through low\-/pressure chemical vapor deposition on an 8-\si{\micro\meter}-thick thermal silicon dioxide layer, mechanically supported by a 50-\si{\micro\meter}-thick silicon substrate. A 4\-/inch wafer was diced in 20~\si{\milli\meter} by 20~\si{\milli\meter} chips, and the integrated circuit was fabricated through e-beam lithography and plasma etching techniques. 

The integrated circuit was patterned onto the chip by spinning a 270-\si{\nano\meter}-thick layer of hydrogen silsesquioxane (HSQ), which was exposed at a dose of 1300~\si{\micro\coulomb\per\centi\meter\squared} and developed in 25\% in tetramethylammonium hydroxide. The pattern was then transferred to the silicon nitride film by inductively coupled plasma (ICP) and reactive ion etching using a $\mathrm{CHF_3/N_2/O_2}$-based chemistry optimized to ensure good sidewall verticality. A 500-\si{\nano\meter}-thick HSQ layer was spun onto the chip and thermally cured to obtain an upper cladding with optical properties close to that of silicon dioxide. The chip was finally cleaved to expose the waveguide facets for the experimental characterization.

\subsection{Experimental Setup}
\begin{figure}[b]
	\centering
	\includegraphics[width = 0.48\textwidth]{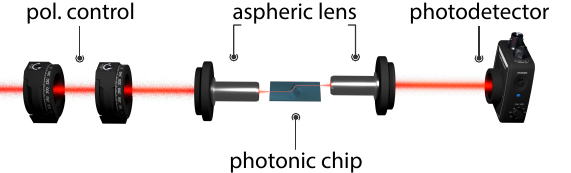}
	\caption{Schematic representation of the experimental setup. See main text for details. 
    }
	\label{fig: setup}
\end{figure}
The integrated devices have been characterized using a chip edge\-/coupling setup, as shown in the schematic of Fig. \ref{fig: setup}.
Various pigtailed laser sources (one at a time) were first collimated using a reflective collimator which produced a \SI{4}{\milli\meter}-diameter beam. Polarization and power control was achieved by combining a broadband half-wave plate and polarizer covering the whole visible spectrum. 
Laser light was coupled into and out of the chip by means of molded aspheric lenses with focal distance $f=3.1$~\si{\milli\meter} and numerical aperture $\mathrm{NA}=0.7$ to ensure a tight focusing spot on the chip facet. The lenses and the chip were respectively aligned via 3-axis and 4-axis micropositioners. 
Output light was finally focused on a variable gain photodetector to measure the output intensity, while a power meter was used to measure the input intensity, allowing to calculate the absolute transmission through the chip. The laser signal was modulated at the input to reliably detect the output intensity via an oscilloscope while minimizing the influence of external noise.

\subsection{Computational Hardware}
All optimizations/simulations were carried out on an NVIDIA RTX 4060 Ti or an NVIDIA RTX 5070 Ti with 8 and 16 GB of VRAM, respectively. The full optimization runtime ranged from 8 to 24 hours depending on the size of the device and simulation domain as well as the required FDTD runtime. We emphasize that all optimizations have been carried out with one simulation per optimized wavelength. In the future, performing simulations with a broadband pulse can drastically reduce the total runtime.

\begin{description}[style=unboxed, leftmargin=0cm, font=\normalfont\bfseries]
    \item[\textbf{Acknowledgments}] We are grateful Yannik Mahlau for most helpful discussions. We thank Flexcompute for providing access to the Tidy3D solver and associated cloud computing resources. The authors acknowledge the technical staff at the James Watt Nanofabrication Centre (JWNC) at the University of Glasgow. \\

    \item[\textbf{Funding}]The authors gratefully acknowledge the financial support provided by the Austrian Federal Ministry of Labour and Economy; the National Foundation for Research, Technology and Development; the Christian Doppler Research Association; and the University of Graz. The financial support by EPSRC (Project No. EP/X012689/1) is gratefully acknowledged. \\
    
    \item[\textbf{Conflict of Interest}] The authors declare no conflicts of interest.
\end{description}


\bibliography{bibliography}

\clearpage

\begin{center}
    \vspace*{3cm}

    {\LARGE\bfseries
    Supplementary Infromation for: \par"Inverse Design of Integrated Photonic Components for Visible\-/Light Applications"
    }

    \vspace{2cm}

\end{center}

\section{\label{sec:exp}Experimental validation of device performance}
This section presents the experimental methodology used to evaluate the individual performance metrics of the inverse-designed photonic components. 

In most cases, the performance of an individual device was characterized using cascaded structures containing different numbers of the inverse-designed component (device). The single-device performance was subsequently extracted by fitting the measured transmission as a function of the number of cascaded devices. Because the transmitted signal decreases with increasing cascade length, measurements involving many devices may approach the experimental noise floor. Such measurements were therefore excluded from the evaluation process to avoid noise-floor-limited data influencing the extracted device performance. Furthermore, for most crosstalk measurements, the experimental signals were already below the experimental noise floor, preventing a reliable experimental quantification of the crosstalk.

\subsection{\label{sec:exp bend} Waveguide Bend}

\begin{figure}[ht!]
    \centering
    \includegraphics[width = 0.45\textwidth]{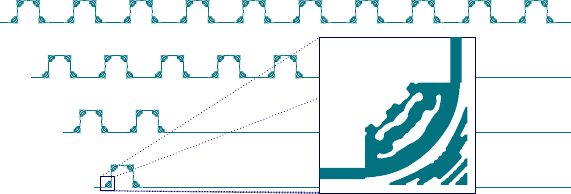}
    \caption[]{Illustration of cascaded waveguide bends with different numbers of elements. Cascades comprising N = 4, 8, 20, and 40 bends are shown. The same arrangement was used to characterize the performance of the conventional bend designs for comparison.}
    \label{fig: sup waveguide bend}
\end{figure}

The performance metric for waveguide bends is their insertion loss, which is determined from the optical transmission. To characterize the performance of both the inverse-designed bend and its conventional counterpart, cascaded test structures containing an increasing number of waveguide bends were fabricated and measured. By evaluating the transmission of cascades with different numbers of bends, as illustrated in Fig. \ref{fig: sup waveguide bend}, the insertion loss of a single bend can be extracted. Specifically, cascades composed of N = 4, 8, 20, 40, 80, 200, 400 bends were characterized to allow a reliable extraction of the insertion loss per bend.

\subsection{\label{sec:exp crossing} Waveguide Crossing}
The performance metrics for waveguide crossings are the insertion loss and crosstalk to the cross ports. The insertion loss is determined from the transmission through the through ports (from l.t.r), whereas the crosstalk is quantified from the power coupled into the cross ports (top/bottom). To characterize both metrics for the inverse-designed and conventional crossing geometries, several measurement configurations were implemented.
\begin{figure}[ht!]
    \centering
    \includegraphics[width = 0.45\textwidth]{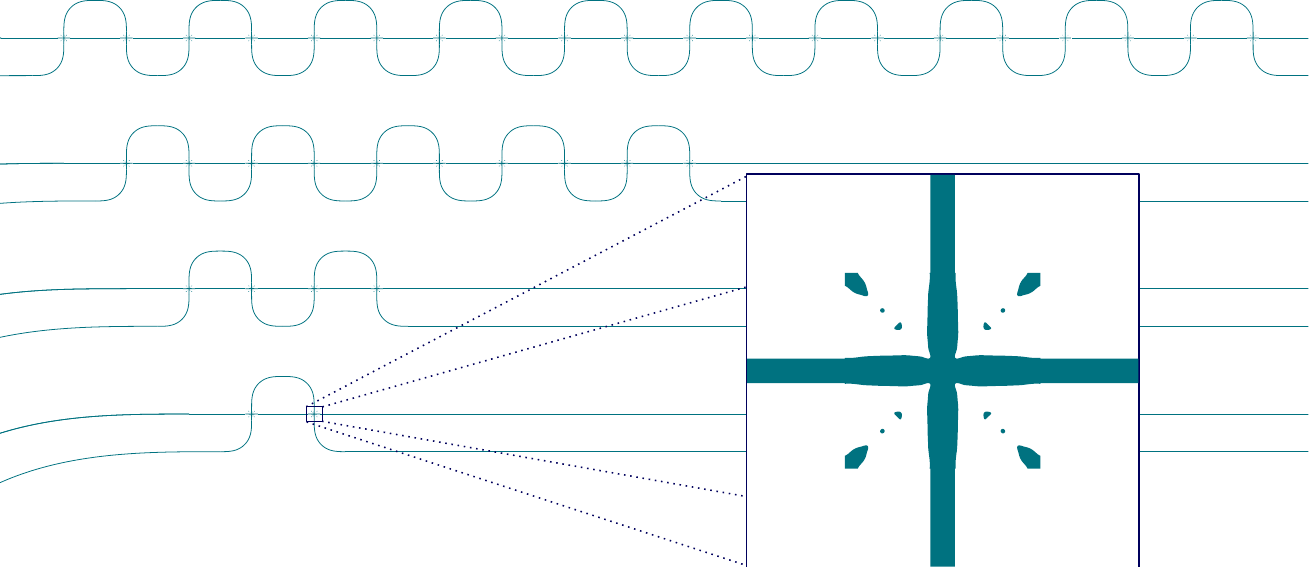}
    \caption[]{Illustration of cascaded waveguide crossings with different numbers of elements. Cascades comprising N = 2, 4, 10, and 20 crossings are shown. A similar arrangement was used to characterize the performance of the conventional crossing design for comparison.}
    \label{fig: waveguide crossings}
\end{figure}

First, cascaded crossing structures with connected cross ports were fabricated, enabling the simultaneous evaluation of insertion loss and crosstalk, as illustrated in Fig.\ref{fig: waveguide crossings}. Cascades containing N = 2, 4, 10, 20, 40, 100 crossings were characterized. However, this configuration may introduce parasitic effects in the transmission and crosstalk measurements due to interference between coupled cross-port modes or resonant effects within the connected waveguide network. Therefore, additional measurements were performed using cascaded crossings with disconnected cross ports to independently evaluate the insertion loss, as well as individual crossing devices with isolated cross ports to directly quantify the intrinsic crosstalk.

\subsection{Power Splitter\label{sec:exp splitter} }
\begin{figure}[ht!]
    \centering
    \includegraphics[width=0.45\textwidth]{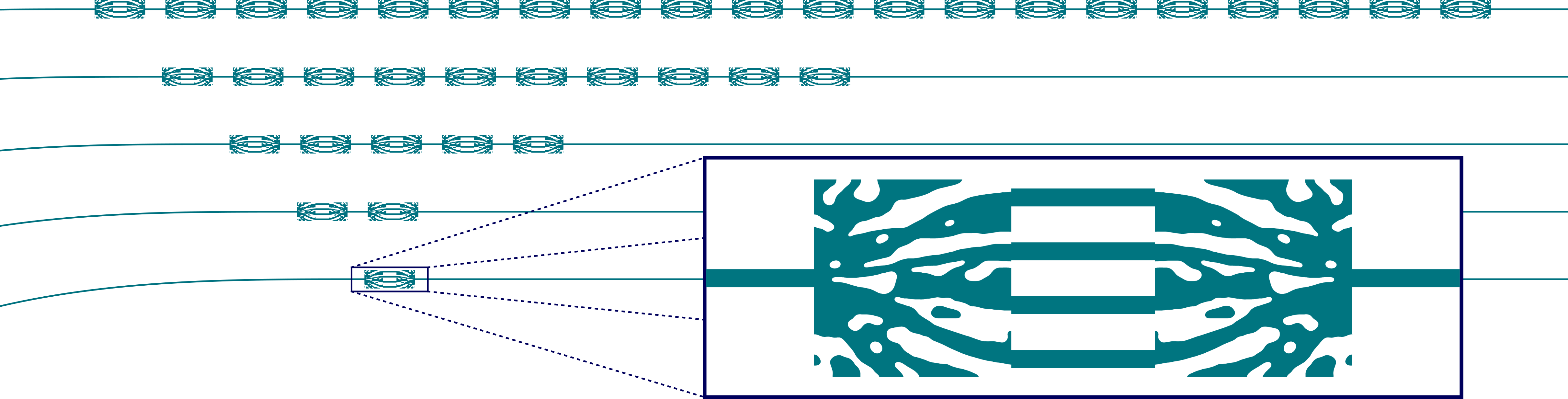}
    \caption[]{Illustration of cascaded power splitters with different numbers of elements. Cascades comprising N = 2, 4, 10, 20, and 40 power splitters are shown. The characterization relies on the reciprocal operation of the power splitter. Under the assumption of reciprocity, every second device is operated in reverse as a power combiner, thereby recombining the optical power from the four output waveguides into a single output waveguide.}
    \label{fig: sup power splitter}
\end{figure}

The performance of the power splitter is evaluated in terms of its insertion loss and output power splitting ratio. As for the previously described components, cascaded devices are employed to determine the insertion loss. A total of N = 2, 4, 10, 20, and 40 splitter stages are characterized. To enable cascade measurements, the reciprocal operation of the device is assumed, and every second splitter is mirrored to operate as a power combiner, thereby recombining the four output waveguides into a single waveguide, as illustrated in Fig. \ref{fig: sup power splitter}.

The output power splitting ratio is evaluated using individual splitter devices. For each device, the optical power at all four output ports is measured and used to determine the splitting ratio. This procedure is repeated for five nominally identical devices to assess the average performance and device\-/to\-/device variation.

\subsection{Mode Converter \label{sec:exp converter} }

\begin{figure}[!ht]
    \centering
    \includegraphics[width=0.45\textwidth]{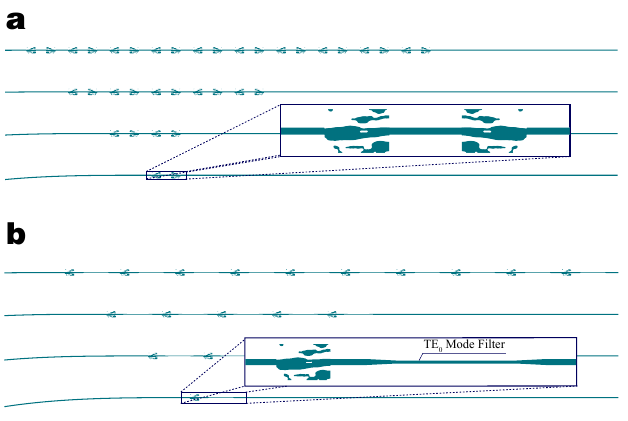}
    \caption[]{(a) Illustration of cascaded mode converters with different numbers of elements. Cascades comprising N = 2, 4, 10, and 20 mode converters are shown. The characterization relies on the reciprocal operation of the mode converter. Under the assumption of reciprocity, every second device is operated in reverse, thereby converting the higher-order TE mode back into the fundamental $\mathrm{TE_0}$ mode.(b) Schematic of the cascaded mode converter measurement structure incorporating single-mode $\mathrm{TE_0}$ filters. The filters reject higher-order modes while transmitting the fundamental $\mathrm{TE_0}$ mode, allowing the residual unconverted power to be quantified.}
    \label{fig: sup mode converter}
\end{figure}

The performance of the mode converter was evaluated in terms of its conversion efficiency and residual power in the unconverted mode. To do so, the insertion loss of the mode converter was evaluated using cascaded devices using a total of N = 2, 4, 10, 20, 100, and 200 converter stages. As for the power splitter, the reciprocal operation of the device was exploited by mirroring every second mode converter, thereby converting the higher\-/order TE mode back to the fundamental $\mathrm{TE_0}$ mode. The corresponding measurement configuration is illustrated in Fig. \ref{fig: sup mode converter} (a).

To quantify the residual power in the unconverted $\mathrm{TE_0}$ mode, a single\-/mode $\mathrm{TE_0}$ filter (reject $\mathrm{TE_1}$) was designed. The filter consits of a narrow waveguide region which slolely supports the fundamental $\mathrm{TE_0}$ mode, thereby suppressing higher\-/order modes. To account for the insertion loss introduced by the filter itself, its transmission was independently characterized using cascaded filter structures. A detailed description of the filter design is shown in section \ref{sec: sm-filter}.

The single\-/mode filters were subsequently employed to isolate the residual TE$_0$ mode at the output of the mode converter. For this purpose, cascaded mode converters incorporating single-mode filters were fabricated and characterized, as illustrated in Fig. \ref{fig: sup mode converter} (b).

\subsection{Mode Devision Multiplexer \label{sec:exp mode splitter} }

The performance of the mode division multiplexer was evaluated in terms of its transmission and crosstalk for the two output ports corresponding to the supported spatial modes. For the transmission measurements, cascaded structures consisting of N = 2, 4, 10, 20 and 40 mode division multiplexers were fabricated and characterized. For the crosstalk measurements, cascades with N = 2, 4, 10, and 20 devices were used, as the increased number of cascaded elements resulted in insufficient signal levels for reliable crosstalk characterization.

First, the overall transmission of the mode division multiplexer was characterized using a fundamental $\mathrm{TE_0}$ mode as the input. Similar to the characterization of the power splitter, the devices were arranged in a cascade configuration. Exploiting the reciprocal operation of the passive devices, every second multiplexer was reversed, thereby recombining the separated $\mathrm{TE_0}$ and $\mathrm{TE_1}$ modes into a single $\mathrm{TE_0}$ waveguide mode.

\begin{figure}[!ht]
    \centering
    \includegraphics[width=0.45\textwidth]{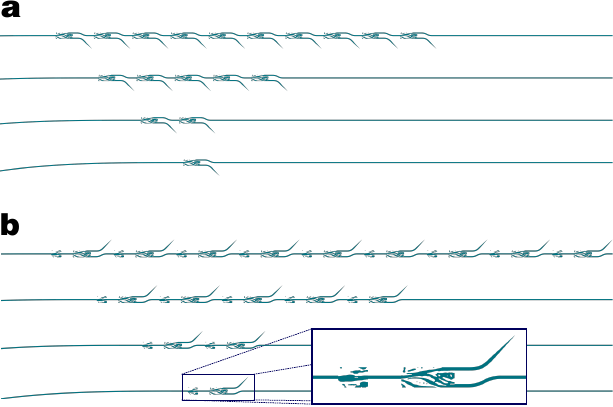}
    \caption[]{(a) Illustration of the cascaded mode sorter measurement structure for characterization of the $\mathrm{TE_0}$ output port. The transmission is characterized using cascades comprising different numbers of mode sorters. (b) Illustration of the cascaded mode sorter measurement structure for characterization of the $\mathrm{TE_1}$ output port. The $\mathrm{TE_1}$ input mode is generated using the $\mathrm{TE_0}$-$\mathrm{TE_1}$ mode converter, enabling characterization of the transmission through the TE$_1$ channel.}
    \label{fig: sup mode sorter}
\end{figure}

The transmission of the $\mathrm{TE_0}$ output port was subsequently characterized using the same cascade length. In this configuration, the $\mathrm{TE_0}$ output port of each multiplexer was connected to the input port of the subsequent device, while the $\mathrm{TE_1}$ port was terminated using a taper structure leading to a waveguide end. This arrangement enables direct characterization of the transmission through the $\mathrm{TE_0}$ channel while suppressing contributions from the $\mathrm{TE_1}$ port. The corresponding measurement configuration is shown in Fig.~\ref{fig: sup mode sorter} (a).

The transmission of the $\mathrm{TE_1}$ output port was evaluated using the previously characterized $\mathrm{TE_0}$\-/$\mathrm{TE_1}$ mode converter. The mode converter was positioned upstream of the input of the multiplexer to excite the $\mathrm{TE_1}$ mode. The $\mathrm{TE_1}$ output ports of consecutive multiplexers were connected, while the $\mathrm{TE_0}$ output ports were terminated, enabling characterization of the $\mathrm{TE_1}$ transmission. The corresponding measurement configuration is shown in Fig.~\ref{fig: sup mode sorter} (b).

The crosstalk performance of the mode division multiplexer was evaluated using analogous cascade configurations. To characterize the crosstalk at the $\mathrm{TE_1}$ output port, a fundamental $\mathrm{TE_0}$ mode was launched into the multiplexer, while the $\mathrm{TE_1}$ output ports were connected to the input ports of the subsequent devices. The measured transmission therefore corresponds to the undesired coupling from the $\mathrm{TE_0}$ input mode into the $\mathrm{TE_1}$ output channel.

Finally, the crosstalk at the $\mathrm{TE_0}$ output port was characterized using a $\mathrm{TE_1}$ input mode generated by the $\mathrm{TE_0}$\-/$\mathrm{TE_1}$ mode converter. In this configuration, the $\mathrm{TE_0}$ output ports were connected to the input ports of the subsequent devices, while the $\mathrm{TE_1}$ output ports were terminated. Consequently, the measured transmission corresponds to the residual power coupled into the $\mathrm{TE_0}$ output channel.

\subsection{Wavelength Division Multiplexer \label{sec:exp color splitter} }

The performance of the wavelength division (de)multiplexer was evaluated in terms of its overall transmission and the relative transmission into the two output ports as a function of wavelength. To characterize the overall transmission, cascaded structures consisting of N = 1, 2, 5, 10, and 20 wavelength division (de)multiplexers were fabricated and measured. The reciprocal operation of the passive device was exploited by operating every second device in a mirrored configuration. In this way, the two separated wavelength channels are recombined into a single waveguide, which is then used as the input for the subsequent device.

The relative transmission into the two output ports was subsequently characterized using five individual wavelength division (de)multiplexers. For each device, the transmission at both output ports was measured over the investigated wavelength range. The measurements were averaged across the five devices to determine the mean relative transmission at each output port.

\subsection{Polarization Splitter \label{sec: exp pol. splitter}}
The polarization splitter was characterized in terms of its transmission and polarization splitting performance. To determine the transmission, cascaded structures containing N = 1, 2, 5, 10, 20, 50, and 100 polarization splitters were fabricated and measured. The single-device transmission was subsequently extracted from the transmission dependence on the number of cascaded devices.

To characterize the polarization splitting, five individual polarization splitters were measured. For each device, the relative transmission at the two output ports was determined for the respective input polarization states. The measurements across the five devices were then averaged to obtain the mean relative transmission.
\nocite{*}

\section{Single Mode Filter Design}
\begin{figure}[ht!]
    \centering
    \includegraphics[width=0.45\textwidth]{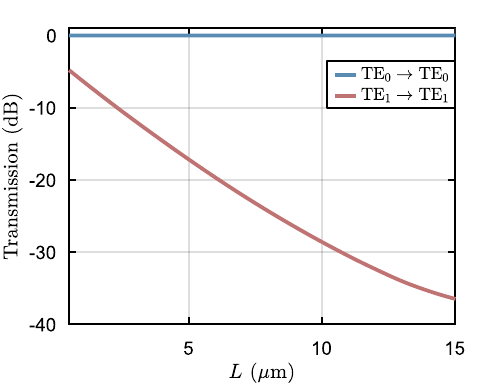}
    \caption[]{Simulated transmission through the single-mode filter as a function of the length $L$ of the 250~\si{\nano\meter} wide waveguide section for the fundamental TE$_0$ and higher-order TE$_1$ modes.}
    \label{fig: transmission sm filter}
\end{figure}

In this section, we describe the functionality of a single-mode filter used to quantify the performance of the inverse-designed mode converter. The filter exploits the fact that sufficiently narrow waveguide sections support only the fundamental waveguide mode while strongly suppressing higher-order modes. The simulated transmission through a 250~\si{\nano\meter} wide waveguide section as a function of its length is shown in Fig.~\ref{fig: transmission sm filter} for both the fundamental TE$_0$ mode and the higher-order TE$_1$ mode. 

For the filter used in the experiments described above, a length of 10~\si{\micro\meter} was selected for the 250~\si{\nano\meter} wide filter section. The narrow section is connected to the 500~\si{\nano\meter} wide waveguides used throughout the photonic platform by two 2.5~\si{\micro\meter} long linear taper regions.

\label{sec: sm-filter}

%

\end{document}